# Next-generation Probabilistic Computing Hardware with 3D MOSAICs, Illusion Scale-up, and Co-design

Tathagata Srimani[1,†], Robert Radway[2], Masoud Mohseni[3], Kerem Çamsarı[4], Subhasish Mitra[5]
[1]Carnegie Mellon University, [2]University of Pennsylvania, [3]Hewlett Packard Labs, [4]UC Santa Barbara, [5]Stanford University, e-mail: tsrimani@andrew.cmu.edu

**Topic (Probabilistic Computing):** The vast majority of 21st century AI workloads are based on gradient-based deterministic algorithms such as backpropagation. One of the key reasons for the dominance of deterministic ML algorithms is the emergence of powerful hardware accelerators (GPU and TPU) that have enabled the wide-scale adoption and implementation of these algorithms. Meanwhile, discrete and probabilistic Monte Carlo algorithms have long been recognized as one of the most successful algorithms in all of computing with a wide range of applications. Specifically, Markov Chain Monte Carlo (MCMC) algorithm families have emerged as the most widely used and effective method for discrete combinatorial optimization and probabilistic sampling problems. We adopt a hardware-centric perspective on probabilistic computing, outlining the challenges and potential future directions to advance this field. We identify two critical research areas: 3D integration using MOSAICs (Monolithic/Stacked/Assembled ICs) and the concept of *Illusion*, a hardware-agnostic distributed computing framework designed to scale probabilistic accelerators.

**Challenges:** Despite their significance in ML and AI, MCMC algorithms have yet to be accelerated with domain-specific hardware and are still primarily run on conventional CPUs or GPUs, severely limiting their widespread adoption. A fundamental challenge in accelerating MCMC algorithms is their inherently *serial* nature which obstructs parallelism. Another practical challenge is the need to generate a massive amount of uncorrelated random numbers (RNG), requiring trillions ($10^{12}$) of them within a few seconds. Even in simplified models like 2D checkerboards implemented on GPUs, sampling throughputs have saturated at around 10 billion ($10^{10}$) RNGs per second per chip, consuming 10 to 100W of power. Exacerbating this problem, today's silicon systems face fundamental limitations: the energy and speed benefits of smaller feature sizes have dramatically slowed over the past decade (*miniaturization wall*), and computing performance and energy efficiency are now dominated by data-movement (*energy/latency*) overheads rather than actual computation (*memory wall*).

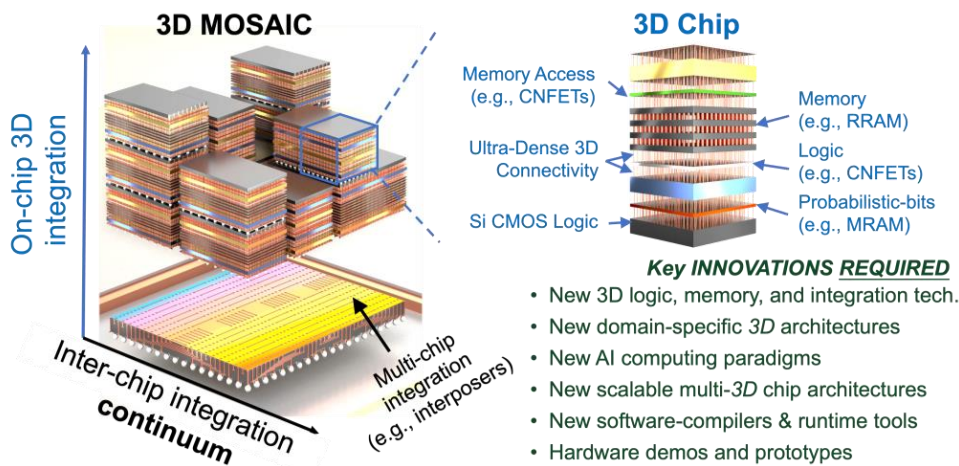

**Fig. 1:** Domain-Specific 3D MOSAICs (Monolithic/Stacked/Assembled ICs) targeting 100-1,000× system-level energy-delay benefits in scientific computing algorithms

**Opportunities:** To overcome these challenges and advance AI-hardware scaling for scientific computing, we need a synergistic combination of new device and integration technologies, circuits, domain-specific architectures, AI-optimized algorithms, compilers, and runtime software support. To advance probabilistic computing hardware, we see two complementary opportunities: one with 3D integration of CMOS + X systems [1], and the other with *Illusion* where a large problem graph can be broken into pieces with minimal communication between interconnected chips [2].

**3D MOSAICS:** We assert that ultra-dense 3D integration of multiple layers of compute and memory access logic will be essential for achieving 100-1,000× improvements in system-level Energy-Delay-Product for scientific computing algorithms (e.g., the powerful Markov Chain Monte Carlo algorithms for combinatorial optimization and sampling) through technology-architecture-software co-exploration on 3D MOSAICs (Monolithic/Stacked/Assembled ICs, Fig. 1). Each 3D chip leverages ultra-dense integration of logic and memory layers to vastly reduce the memory-to-compute data movement overheads, resulting in substantial system-level energy and throughput benefits. Multiple 3D chips can be integrated using a combination of chip stacking, interposer, and wafer-level assembly techniques. In 3D MOSAICs, benefits can be expected at various levels of the stack:

1. **Device and Integration Level**: New materials enhance functionality (e.g., carbon nanotube FETs, RRAM/MRAM integrated on top of silicon CMOS).
2. **Circuit and Architecture Level**: 3D integration can enable new communication capabilities across chips, leading to large-scale probabilistic computation with millions of probabilistic bits.
3. **System and Software Level**: New software compilation and runtime technologies can complement domain-specific multi-chip architectures and enable efficient software execution on 3D MOSAICs.

*Illusion* **for probabilistic computing**: Beyond 3D MOSAICs, another effective method in scaling up computation is Illusion. The concept of Illusion was initially demonstrated in DNN inference to address limited on-chip memory, which necessitated frequent and costly off-chip memory accesses. By networking multiple chips, each with a minimal amount of local memory and rapid wakeup/shutdown techniques, Illusion achieved energy and execution times close to an ideal single-chip solution without off-chip memory.

We believe that this concept can be adapted to probabilistic computing [3] in a *hardware-agnostic* manner. This would involve breaking a large problem graph into smaller pieces through graph partitioning – potentially using weighted min-cut algorithms – and distributing the graph across interconnected chips. Thanks to the forgiving nature of probabilistic algorithms in sampling and optimization, the sampling throughput and the accuracy of an *ideal (*hypothetical) probabilistic computer that can house the entire graph can closely be approximated in synchronous and asynchronous architectures. Importantly, the concept of Illusion is *agnostic* to the choice of a probabilistic accelerator whether it is based on MRAM-based probabilistic bits, coupled oscillators, or other Ising machines.

*Outlook:* The main challenges in probabilistic computing with domain-specific hardware are related to scaling up the nodes, increasing the number of parameters and their interaction (2nd order, higher order, etc.) in a graph, and maintaining a very large throughput. We believe that MOSAIC3D and Illusion approaches are two complementary methods that enable the large-scale deployment of probabilistic computers, with a co-design approach across the stack.